\documentclass[aps,prb,twocolumn,showpacs,superscriptaddress]{revtex4-1}

\usepackage{amsmath,amssymb,graphicx,color,float}

\newcommand{\eq}[1]{(\ref{eq:#1})}
\newcommand{\re}{\textrm{Re}}

\begin{document}

\title{Energy Relaxation in one-dimensional polariton condensates}

\author{M. Wouters}
\author{T. C. H. Liew}
\author{V. Savona}
\affiliation{Institute of Theoretical Physics, Ecole Polytechnique F\'{e}d\'{e}rale de Lausanne EPFL, CH-1015 Lausanne, Switzerland}

\date{\today}

\begin{abstract}
We study the kinetics of polariton condensation accounting for the condensation process as well as the energy relaxation of condensed polaritons due to their scattering with phonons and excitons. By assuming a Boltzmann kinetic description of the scattering process, we show that intra-condensate relaxation can be accounted for by an additional time-dependent term in the Gross-Pitaevskii equation. As an example, we apply the formalism to the experimental results recently obtained in polariton microwires [E. Wertz, et al., Nature Phys. {\bf 6}, 860 (2010)~\cite{Wertz2010}]. In the presence of a local non-resonant optical pump, a dynamic balance between spatially dependent relaxation and particle loss develops and excites a series of modes, roughly equally spaced in energy. Upon comparison, excellent agreement is found with the experimental data.
\end{abstract}


\maketitle

\section{Introduction}

Researchers now have two decades of experience in the study of semiconductor microcavities~\cite{Kavokin2007} and exciton-polaritons, the mixed quasiparticles that arise from strong light-matter coupling. Inspired  by the early prediction of Bose-Einstein condensation of excitons~\cite{Moskalenko1962,Blatt1962,Keldyh1968}, impressive progress in cold-atom condensation~\cite{Bloch2008}, and motivated by the prediction of ultra-low threshold polariton lasing~\cite{Imamoglu1996,Malpuech2002}, many experiments have been dedicated to the realization of a Bose-Einstein condensate of polaritons~\cite{Kasprzak2006,Balili2007,Lai2007,Deng2007}. As mixed quasiparticles, polaritons are expected to interact  with each other (due to underlying exciton-exciton interactions) whilst maintaining a light effective mass (due to the trait of cavity photons). As bosons, polaritons should relax into the ground state of the system, producing a transition to a phase coherent state capable of a laser-like emission without the need for population inversion~\cite{Christopoulos2007,Bajoni2008a,Christmann2008,Forrest2010}. This transition is the non-equilibrium analog of Bose-Einstein condensation and can occur at high temperature due to the light effective polariton mass. Although research studies usually use an optical excitation of polaritons, progress in electrical injection sets a direct path for practical applications in the form of polariton-based light-emitting diodes~\cite{Tsintzos2008,Bajoni2008b,Tsintzos2009}. A growing trend towards studying highly confined structures, in which polaritons are confined into zero or one dimensional structures~\cite{Bajoni2007,IdrissiKaitouni2006,Daif2006,Bajoni2008a,Cerna2009}, opens the way to the realisation of microstructured polariton devices.

A theoretical description of the condensation of microcavity polaritons is a challenging~\cite{Keeling2007,Sarchi2007,Sarchi2008,Deng2010} task, especially when one wishes to account for spatial dependence. Since the polariton lifetime is shorter than the relaxation time (in today's samples), the system always lies in a state out of thermal equilibrium, with no well defined chemical potential or temperature. For this reason one requires a kinetic model that accounts for the excitation, scattering, relaxation and decay processes occurring in the system. Semiclassical Boltzmann rate equations~\cite{Tassone1997} can be used to describe the thermal relaxation processes that lead to condensation, however they completely lack of a description of the collective excitations that characterize a quantum fluid. To correctly describe these properties at thermal equilibrium, symmetry-breaking theories are instead necessary. These approaches model the condensate as a classical field, in terms of the Gross-Pitaevskii equation, possibly coupled to a quantum model for the excited states, in analogy to the theory of atomic condensates~\cite{Shi1998,Keeling2007,Sarchi2008}.  Intuition suggests that we can use a mix of the two theories in which higher energy excitons are described by Boltzmann-like equations and condensed polaritons by Gross-Pitaevskii-like equations. Such a model was constructed by Wouters and Carusotto~\cite{Wouters2007} to describe polariton condensates, whilst similar approaches have been developed in atomic condensates~\cite{Gardiner1998a,Gardiner1998b,Zaremba1999}.

Still, in experiments condensation can take place into a state above the ground state~\cite{Krizhanovskii2009,Maragkou2010} and in planar microcavities condensed polariton relaxation was evidenced down a potential gradient~\cite{Balili2007}. In periodic structures, condensation was observed into both in-phase and antiphase Bloch states and the further relaxation into the lower energy in-phase state demonstrated with increasing pump intensity~\cite{Lai2007}. Particularly stunning relaxation effects have appeared in recent experiments on polariton microwires~\cite{Wertz2010}. Here condensation was observed into a series of energy modes that appear roughly equally spaced in energy. Surprisingly, the energy spacing is too large to be attributed to quantization of the center-of-mass motion (the microwire is too long). To describe these effects theoretically it is essential to have a model that accounts for both the energy relaxation of excitons into condensed polaritons and relaxation of polaritons {\it within} the condensed fraction of the polariton gas (polaritons can relax their energy by scattering with phonons or hot excitons). While the first order process is included in most kinetic theories of polariton condensation, the second has never been theoretically addressed, to our knowledge.

In this work we derive a generalized version of the Gross-Pitaevskii equation (GPE), that describes intra-condensate relaxation induced by the coupling with a reservoir. The starting point of the derivation is the Boltzmann collision term for the condensate density. As a result, the GPE acquires an additional time-dependent rate that models the relaxation process. As an example, we apply the formalism to the recent experiment carried out by Wertz {\em et al.} on polariton microwires.\cite{Wertz2010}. The incoherent optical pump is modeled using a classical rate equation coupled to the generalized GPE. The optical pump itself introduces a spatially structured potential (an effect also shown in experiments using resonant excitation~\cite{Amo2010}). Only a full account of the energy relaxation within the condensate and of the spatial dynamics of polaritons interacting with this potential can reproduce the various modal patterns measured experimentally.

The paper is organized as follows. In Sec.~\ref{sec:relax}, we derive the term in the GPE that describes the relaxation. Simulations of relaxation in nonresonantly excited 1D wires are presented in Sec.~\ref{sec:sim}. Conclusions are presented in Sec.~\ref{sec:conclusions}.

\section{Description of the relaxation \label{sec:relax}}

The degeneracy of the Bose gas implies that a good account of the condensate kinetics is already given by a mean field description, formulated in terms of a classical field $\psi(x,t)$. This approach is common to most recent theoretical works~\cite{Malpuech2007,Solnyshkov2008,Keeling2008,Solnyshkov2009,Fraser2009,Shelykh2009,Borgh2010,TralleroGiner2010} that describe most existing experimental data obtained above the condensation threshold. More advanced quantum kinetic theories are only needed to model the details of phase fluctuations~\cite{Carusotto2005,Read2009,Wouters2009}.

A kinetic model formulated in terms of transitions between states characterized by their frequencies, involves a coarse graining of the time. This is analogous to the coarse graining of space used in the semiclassical Boltzmann equation~\cite{boltzmann} to describe the density as a function of position and momentum while fulfilling the constraint imposed by the uncertainty principle. We will use the following  definition of the time dependent spectrum of the bose field $\psi(x,t)$:
\begin{equation}
\psi_\omega(x,t)=\frac{1}{T}\int_{t-T}^t e^{i\omega t'}\;\psi(x,t') dt',
\label{eq:ft}
\end{equation}
where $T$ is the coarse graining time step. When using this formalism to model a steady-state spectrum, it is important that $\hbar/T$ is chosen smaller than the energy spacing of the features to be resolved. Then, the physical result should not be affected by the precise value of $T$. The inverse transform reads
\begin{equation}
\psi(x,t)=\sum_\omega e^{-i\omega t} \psi_\omega(x,t).
\label{eq:ift}
\end{equation}
Spontaneous scattering is a quantum feature that cannot be described within a classical field model such as the GPE and has only been treated in a handful of theories~\cite{Mieck2002,Wouters2009,Read2009}. Neglecting spontaneous scattering might limit the accuracy in modeling the kinetics of low-occupation states. It is however a particularly well suited assumption when modeling intra-condensate kinetics. Indeed, due to the high occupation of both initial and final states in a scattering process, stimulated scattering is largely dominant. We therefore restrict to the stimulated relaxation processes. At high densities, relevant for polariton condensates, these processes are mainly polariton-exciton scattering~\cite{Porras2002} rather than polariton-phonon scattering~\cite{Tassone1997}. The in-scattering term at frequency $\omega$ reads
\begin{equation}
dn_\omega(x,t) = \, n_\omega(x,t)\, \sum_{\omega'} r(\omega,\omega') n_{\omega'}(x,t)\;dt,
\label{eq:dnomega}
\end{equation}
where $r(\omega,\omega')$ is the net scattering rate from the mode at frequency $\omega'$ to the mode at $\omega$. Due to particle number conservation, the relaxation rate obeys the relation $r(\omega,\omega')=-r(\omega',\omega)$.
The first term in the expansion of the rate as a function of the frequency difference is therefore
\begin{equation}
r(\omega,\omega')=\kappa \,\times\, (\omega'-\omega)+\mathcal{O}\left((\omega'-\omega)^3\right),
\label{eq:rw}
\end{equation}
where the relaxation constant $\kappa$ has the dimension of an inverse density. Here, we don't use a microscopic model of the scattering process. Rather, we keep $\kappa$ as a free parameter that we adjust to fit the experimental data. The resulting value is consistent with the scattering on hot excitons as discussed by Porras, et al.~\cite{Porras2002}.

Under the change of the density~\eq{dnomega}, the wave function varies as
\begin{equation}
\psi_\omega(x,t)+d\psi_\omega(x,t)= \sqrt{\frac{n_\omega(x,t)+dn_\omega(x,t)}{n_\omega(x,t)}} \psi_\omega(x,t),
\label{eq:dpsiomega}
\end{equation}
where we used the fact that a stimulated relaxation process does not change the phase of $\psi$. Expanding Eq.~\eq{dpsiomega} to first order in $dn_\omega$, substituting Eqs.~\eq{dnomega} and~\eq{rw} and using the inverse transform~\eq{ift},
one obtains for the relaxation dynamics of the wave function
\begin{equation}
\frac{\partial\psi(x,t)}{\partial t} = \frac{\kappa \bar{n}(x,t)}{2}\left[\bar{\mu}(x,t)-\frac{i \partial}{\partial t}
\right] \psi(x,t).
\label{eq:dpsi1}
\end{equation}
In the derivation, boundary terms in the partial integration that scale as $1/T$ were neglected, because we consider the limit of a large coarse graining time where $1/T$ is smaller than any other relevant frequency scale. In Eq.~\eq{dpsi1}, $\bar{n}$ and $\bar{\mu}$ are defined as
\begin{eqnarray}
\bar{n}(x,t)&=&\frac{1}{T}\int_{t-T}^t |\psi(x,t')|^2 dt',\label{eq:Nb} \\
\bar{\mu}(x,t)&=&\frac{1}{\bar{n}(x,t)}\re\left[\frac{1}{T}\int_{t-T}^t \psi^*(x,t')
 \frac{i\partial}{\partial t} \psi(x,t') dt'\right].
\label{eq:Eb}
\end{eqnarray}
The first expression defines the density averaged over the coarse graining interval $T$. Analogously, $\hbar\bar{\mu}$ can be suggestively read as an averaged chemical potential. We stress however that no actual chemical potential can be defined for this non-equilibrium system, and Eq.~(\ref{eq:Eb}) draws its only justification from the formal derivation presented above. The real part is taken in Eq.~\eq{Eb}, because it is readily shown with the Madelung transformation $\psi=\sqrt{n}\,e^{i\theta}$, that the imaginary part of the integral in Eq.~\eq{Eb} scales as $1/T$. Equation~\eq{dpsi1} gives the term that was sought to describe the stimulated relaxation of a classical Bose field due to local scattering with phonon like particles and is the central result of this work.

The right hand side in Eq.~\eq{dpsi1} resembles the frequency dependent amplification term that was introduced in Ref.~\onlinecite{nonres_superfl} to describe an energy dependent gain mechanism due to scattering from a reservoir: $\partial \psi/\partial t=(P/\Omega_K)[\Omega_K-i\partial/\partial t] \psi$. Two main differences should be emphasized. First, here the relaxation term is proportional to the polariton density $\bar n$, whereas in Ref.~\onlinecite{nonres_superfl}, the relaxation is proportional to the gain from the reservoir $P$. The second important difference is that the gain cutoff frequency $\Omega_K$ is replaced by the average frequency $\bar \mu$. As a consequence, particle loss and gain balance each other and there is no net gain in Eq.~\eq{dpsi1}. The modes with frequency $\omega>\bar \mu$ experience loss, where the ones with $\omega<\bar \mu$ are amplified. Energy absorption from the reservoir into the condensate is thus not included in our model. The relaxation~\eq{dpsi1} is due to the interaction with an environment that is effectively at zero temperature.

\section{Relaxation in 1D wires \label{sec:sim}}

As an example of application, we consider a 1D microwire~\cite{Wertz2010} under non-resonant optical excitation. In most situations in which an incoherent optical pump is present, a good description of the condensation kinetics is given by coupling the GPE to a kinetic equation for the reservoir population.\cite{Wouters2007} To this purpose, we introduce a complex condensate field, $\psi(x,t)$, coupled to a reservoir population, $N(x,t)$. Neglecting the polarization degree of freedom, the evolution of condensed polaritons is given by the generalized GPE for a Bose gas with losses and gain from a reservoir, supplemented with the dissipation term \eq{dpsi1}:
\begin{align}
i\hbar\frac{\partial \psi(x,t)}{d t}&=\left(\hat{E}_{LP}+\frac{i\hbar\left[R_R N(x,t)-\Gamma_C\right]}{2}\right.\notag\\
&\hspace{5mm}\left.+V(x,t)+g\left|\psi(x,t)\right|^2\right.\notag\\
&\hspace{5mm}\left.+\frac{\hbar\kappa\bar{n}(x,t)}{2}\left[i\bar{\mu}(x,t)+\frac{\partial}{d t}\right]\right)\psi(x,t),\label{eq:GP}
\end{align}
where the kinetic energy eigenvalues of $\hat{E}_{LP}$ are:
\begin{equation}
E_{LP}(k)=\frac{E_C(k)+E_X(k)}{2}-\frac{1}{2}\sqrt{\left(E_C(k)-E_X(k)\right)^2+\Omega^2}.\label{eq:Disp}
\end{equation}
Here $E_C(k)$ and $E_X(k)$ represent the cavity photon and exciton dispersion relations, respectively, and $\Omega$ is the exciton-photon coupling constant. Polaritons scatter from the reservoir into the condensate at a rate $R_R$ and radiatively decay at a rate $\Gamma_C$. $V(x,t)=\hbar R_R N(x,t)+\hbar G P(x)$ is an effective potential experienced by polaritons, caused by interactions between polaritons with the reservoir and also the pump field, $P(x)$, described with a strength $G$. The large potential experienced by polaritons at the ends of the wire is accounted for through the introduction of boundary conditions $\psi(x=\pm L)=0$.

The dynamics of the reservoir field, representing optically injected hot excitons injected, is described by:
\begin{equation}
\frac{\partial N(x,t)}{\partial t}=-\left(\Gamma_R+\beta R_R\left|\psi(x,t)\right|^2\right)N(x,t)+P(x)~\label{eq:Reservoir}
\end{equation}
where $\Gamma_R$ is the decay rate of reservoir polaritons and the parameter $\beta<1$ quantifies the backaction of the condensate on the reservoir~\cite{Wouters2009}. Such a parameter is necessary due to the simplified model of a single reservoir used for the exciton states; if $\beta=1$, then unphysical instabilities may develop in the solution of the coupled Eqs.~(\ref{eq:GP}) and (\ref{eq:Reservoir})~\cite{Wouters2009}. The reason is that for $\beta=1$ the gain saturation mechanism is overestimated, given the simplified description of the reservoir in terms of Eq.~(\ref{eq:Reservoir}). At the same time, the assumption $\beta=0$ would make the condensate density diverge, because the gain saturation mechanism is needed to determine the steady state condensate density at a given pump intensity. Ultimately, a microscopic description of the reservoir made with a continuous spectrum of modes, would in principle make the introduction of this phenomenological parameter unnecessary. The choice $\beta<1$ is therefore obliged, given the reservoir model (\ref{eq:Reservoir}). Finally, note that the value of $\beta$ makes no qualitative difference to the spectral and spatial condensate pattern (implying only an overall dependence of the condensate density) so long as it is small enough to prevent instabilities.

We have solved Eqs.~(\ref{eq:GP}-\ref{eq:Reservoir}) with parameters~\cite{Parameters} corresponding to the experiment of Wertz~\cite{Wertz2010} et al., who considered two cases: excitation at the wire center and excitation near one edge. In both cases the time evolution is calculated over a long ($4000$ps) period during which the system reaches a steady state. Energy distributions are then calculated by Fourier transforming the fields.

\subsection{Excitation in wire center}

\begin{figure}[h]
\centering
\includegraphics[width=8.116cm]{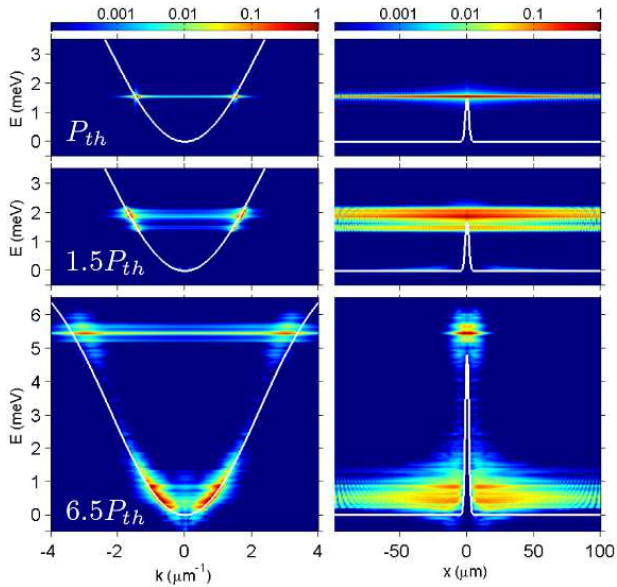}
\caption{(color online) For excitation in the wire center, the energy distribution of condensed polaritons in real (right) and momentum (left) space is shown for increasing pump powers. The white curves show the bare polariton dispersion given by Eq.~(\ref{eq:Disp}) in momentum space (left) and the effective polariton potential, $V(x)$, in real space (right). In each plot the intensity is scaled to the maximum intensity and plotted with a logarithmic color scale. The data were convoluted with a small spectral linewidth ($\delta=0.02$meV) to account for the finite spectral resolution in the experiment.}
\label{fig:Ecenter}
\end{figure}

The energy distributions with excitation at the wire center are shown in reciprocal (left) and real space (right) in Fig.~\ref{fig:Ecenter}. The presence of the pump introduces excitons into the reservoir and also creates the spatially varying potential, $V(x)$. Together with a slight blueshift caused by polariton-polariton interactions, this potential determines the energy at which a polariton condensate forms due to the scattering of particles from the reservoir. The condensed particles are created with zero in-plane wavevector, in the center of the pumping spot, and are pushed away by the potential $V(x)$ so that they fill the wire. As the pump intensity increases, a clear blueshift of the condensed polaritons is visible and stimulation of the energy relaxation takes over. Propagating polaritons form a spectral pattern with several modes, roughly equally spaced, in agreement with the experimental results~\cite{Wertz2010}.\\

\begin{figure}[h]
\centering
\includegraphics[width=8.116cm]{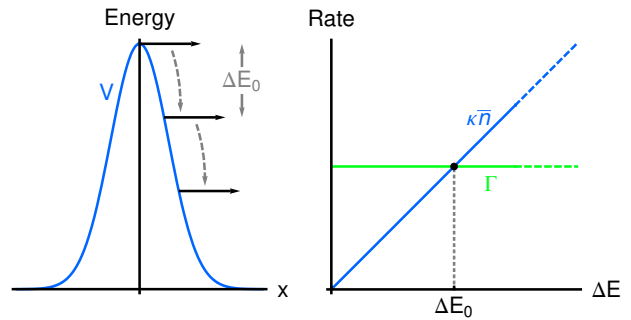}
\caption{(color online) a) Illustration of quantized energy relaxation as polaritons propagate down the pump-induced potential, $V(x)$. b) Sketch of the incoming (proportional to $\kappa \bar{n}(x,t)$) and outgoing rates ($\Gamma$) for the lower energy mode in a relaxation step, as a function of the energy difference with a higher energy mode.}
\label{fig:Erelaxation}
\end{figure}

The energy relaxation introduces a gain mechanism for the lower energy polariton states and one can imagine the relaxation of polaritons as they travel down the pump-induced potential, as illustrated in Fig.~\ref{fig:Erelaxation}a. The rate of relaxation between any two modes is an increasing function of their energy difference, $\Delta E$,  to lowest order according to Eq.~(\ref{eq:rw}), as illustrated by the blue (dark-grey) line in Fig.~\ref{fig:Erelaxation}b. For the formation of the condensate to occur in a lower energy mode, in the steady state, the gain must exactly balance the losses in the process. In other words, the incoming rate due to this relaxation must balance the outgoing rate, which is mostly due to radiative decay, illustrated by the green (light-grey) line in Fig.~\ref{fig:Erelaxation}b. The balance occurs at the energy $\Delta E_0$ and relaxation is only expected to appear between modes separated at this energy. The further outward propagation of polaritons can allow a cascade effect in which polaritons relax in units of $\Delta E_0$ into a series of levels (Fig.~\ref{fig:Erelaxation}a). As the rate is proportional to $\kappa \bar{n}(x,t)$ and to the energy step, the modes are roughly equally spaced in energy if we assume that the density in the initial mode is approximately the same for all relaxation steps. The data by Wertz et al. support this.

\begin{figure}[h]
\centering
\includegraphics[width=8.116cm]{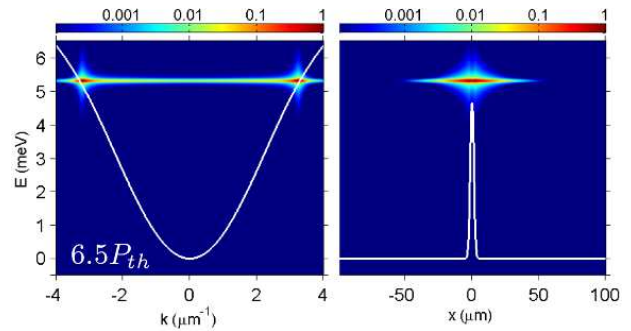}
\caption{(color online) Same as the bottom panels of Fig.~\ref{fig:Ecenter} with $\kappa=0$.}
\label{fig:kappa0}
\end{figure}

To confirm the necessary role of energy relaxation in the generation of a series of condensate modes at discrete energies, we repeated our calculations setting $\kappa=0$. Fig.~\ref{fig:kappa0} shows that in this case all polaritons condense into a single energy mode determined mostly by $V(x=0)$.

\subsection{Excitation near wire edge}

\begin{figure}[ht]
\centering
\includegraphics[width=8.116cm]{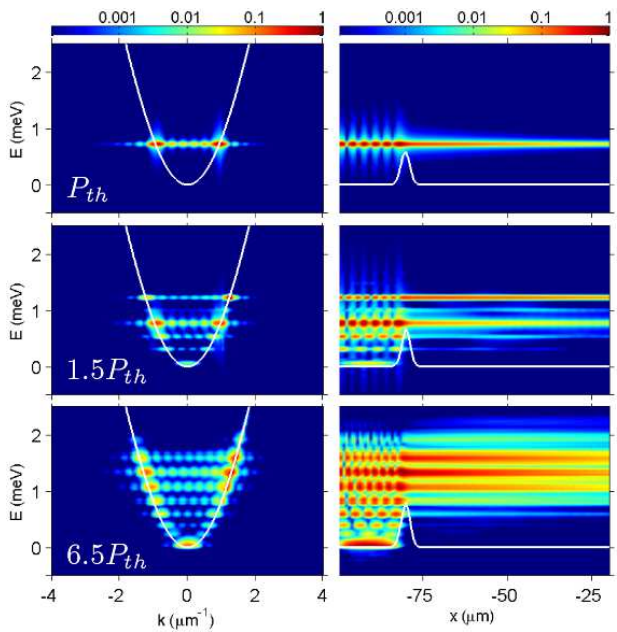}
\caption{(color online) Same as in Fig.~\ref{fig:Ecenter} for excitation near the wire edge.}
\label{fig:Eedge}
\end{figure}

The case of excitation near the wire edge is shown in Fig.~\ref{fig:Eedge}. The same phenomenology as in the case of excitation at the wire center can be identified; condensation at a blueshifted energy and energy relaxation at higher intensities. However, one can now observe the striking appearance of harmonic modes due to a potential trap formed between the wire edge and the pump-induced potential, $V(x)$. This feature is in close agreement with the experimental results~\cite{Wertz2010}. The positioning of the optical pump near the wire edge introduces a strong asymmetry between relaxation on the two sides of the pump. It is observed theoretically and experimentally that relaxation is more effective when polaritons interact with the wire edge.

We note that our model is limited in its quantitative description of the power dependence. Particularly near the wire edge, higher pump powers were required experimentally to observe the same effects. Our energy relaxation model assumes that the relaxation rate is proportional to the condensed polariton population and the energy difference in a relaxation process. In reality, we expect the rate to also depend on the total exciton population and not to be such a simple function of the energy difference. The study of these issues is beyond the scope of our work; it would require further experimental characterization and a microscopic model of the polariton-reservoir kinetics. Although we also neglected exciton diffusion, intrinsic reservoir dynamics and nonlinear loss mechanisms, we believe that our model captures the most important elements of the physics involved and reproduces the experimental character.

\section{Conclusions \label{sec:conclusions}}

We have presented a theoretical formalism that models both the condensation process, as polaritons relax from high-energy hot-exciton states into the condensed polariton field, as well as further energy relaxation taking place within the condensate itself as polaritons interact with excitons and phonons. The theory goes beyond the standard Gross-Pitaevskii theories, which do not describe the condensation process or energy relaxation, and the Boltzmann rate equations, which do not allow a description of coherence, interference or the structure of modes in confined systems.

We applied the formalism to the recent experiment of Wertz~\cite{Wertz2010} et al. This experiment uncovered a number of remarkable effects in the physics of nonequilibrium Bose-Einstein condensation in confined systems. The presence of optically induced potentials changes the modal distribution in the system. Condensation initiates strongest locally, at the position of the optical pumping. However, the dispersion of polaritons and their further relaxation allows lower energy modes to host the condensate in a spatially dependent way. Our model reproduces well the experimental findings. In particular it explains the puzzling feature of discrete energy modes of the condensate, observed experimentally. The excitation of this mode ladder is naturally attributed to the energy relaxation mechanism within the condensate. A selection rule in the energy relaxation steps is determined by the balance between gain and loss for the final energy mode of the relaxation process. In the case of excitation at the wire edge, the narrow laser-induced trap creates a clear set of harmonic modes.

We aknowledge enlightening discussions with J. Keeling and A. Kavokin. This work was supported by NCCR Quantum Photonics (NCCR QP), research instrument of the Swiss National Science Foundation (SNSF).

\end{document}